\documentclass[aps,prd,preprint,nofootinbib,11pt]{revtex4}
\usepackage{amsfonts}
\usepackage{mathrsfs}
\usepackage{graphicx}
\usepackage{amsmath}
\usepackage{amssymb}
\usepackage{subfigure}
\usepackage{epsfig}
\usepackage{graphicx}
\usepackage{ulem}
\usepackage{color}
\newcommand{\bqa}{\begin{eqnarray}}
\newcommand{\eqa}{\end{eqnarray}}
\newcommand{\beq}{\begin{equation}}
\newcommand{\eeq}{\end{equation}}
\allowdisplaybreaks[1]
\graphicspath{{fig/}{dia/}} \DeclareGraphicsExtensions{.eps}

\begin{document}
\title{\Large Hidden-charm and -bottom tetraquark states with $J^{PC}=1^{-+}$ via QCD sum rules\\[7mm]}

\author{Bing-Dong Wan$^{1,2}$\footnote{wanbd@lnnu.edu.cn}, Yan Zhang$^{1,2}$, Jun-Hao Zhang$^{1,2}$ and Ming-Yang Yuan$^{1,2}$ \vspace{+3pt}}

\affiliation{$^1$Department of Physics, Liaoning Normal University, Dalian 116029, China\\
$^2$ Center for Theoretical and Experimental High Energy Physics, Liaoning Normal University, Dalian 116029, China}

\author{~\\~\\}

\begin{abstract}
\vspace{0.3cm}
We investigate the $1^{-+}$ hidden-charm and hidden-bottom tetraquark states within the framework of QCD sum rules. The mass spectra are computed by including condensates up to dimension eight in the operator product expansion. Our results indicate the possible existence of four $1^{-+}$ hidden-charm tetraquark states, with predicted masses of 
$(4.83 \pm 0.15)$ GeV, $(4.88 \pm 0.18)$ GeV, $(4.72 \pm 0.16)$ GeV, and $(4.79 \pm 0.12)$ GeV, while their hidden-bottom counterparts are estimated to have masses of 
$(11.08 \pm 0.16)$ GeV, $(11.16 \pm 0.14)$ GeV, $(10.99 \pm 0.16)$ GeV, and $(11.03 \pm 0.15)$ GeV, respectively. We also analyze the possible decay modes of these tetraquark states, which may be accessible in future experiments at BESIII, Belle~II, LHCb, and future STCF. These findings provide valuable guidance for the experimental search for exotic $1^{-+}$ tetraquark states in both the charm and bottom sectors.
 \end{abstract}
 \maketitle
\newpage

\section{Introduction}

Since the discovery of the $X(3872)$~\cite{Choi:2003ue}, significant progress has been made in the experimental exploration of heavy exotic hadrons. A multitude of unconventional states, collectively referred to as the $XYZ$ family, have been observed in the charmonium and bottomonium regions by Belle, BESIII, LHCb, and other collaborations~\cite{Brambilla:2019esw,Chen:2016qju,Liu:2019zoy,Guo:2017jvc,Wang:2025sic}. These structures cannot be easily accommodated within the conventional quark model, in which mesons are quark--antiquark pairs and baryons consist of three quarks, and their properties point towards more complex configurations, including tetraquarks, hadronic molecules, hybrids, and other multiquark systems.

Particular attention has been devoted to states with exotic quantum numbers, such as $0^{--}$, $0^{+-}$, $1^{-+}$, and $2^{+-}$, which are forbidden in conventional quark--antiquark or three-quark configurations. Such states cannot mix with ordinary hadrons and therefore provide a clean probe of genuinely exotic structures and the nonperturbative dynamics of QCD. Experimental evidence, including the observation of the $\eta_1(1855)$ by BESIII~\cite{BESIII:2022riz,BESIII:2022iwi}, confirms the existence of mesons with exotic quantum numbers and motivates further searches in both charm and bottom sectors.

Among these exotic states, tetraquark candidates with hidden heavy flavors, with quark content $Q\bar{Q}q\bar{q}$ ($Q=c,b$ and $q=u,d$), have attracted particular interest. The interplay between heavy and light quarks provides a unique laboratory to study nonperturbative QCD dynamics. Observed charged states in the $Z_c$ and $Z_b$ mass regions indicate the presence of at least four constituent quarks, supporting the tetraquark interpretation. In particular, states with $J^{PC}=1^{-+}$ are theoretically intriguing, as they can arise from diquark--antidiquark configurations or meson--meson molecular structures, yet different models predict distinct masses and decay patterns.

The QCD sum rule (QCDSR) approach offers a first-principle framework to investigate such states~\cite{Shifman}, and this methodology has been extensively and successfully applied to the study of both conventional and exotic hadrons~\cite{Albuquerque:2013ija,Wang:2013vex,Govaerts:1984hc,Reinders:1984sr,P.Col,Narison:1989aq,Tang:2021zti,Qiao:2014vva,Qiao:2015iea,Tang:2019nwv,Wan:2020oxt,Wan:2022xkx,Zhang:2022obn,Wan:2024dmi,Tang:2024zvf,Li:2024ctd,Zhao:2023imq,Yin:2021cbb,Yang:2020wkh,Wan:2024cpc,Wan:2024pet,Wan:2024ykm,Zhang:2024jvv,Tang:2024kmh,Tang:2016pcf,Tang:2015twt,Qiao:2013dda,Qiao:2013raa,Wan:2020fsk,Wan:2025xhf,Chen:2014vha,Azizi:2019xla,Wang:2017sto,Chung:1981wm,Wan:2025fyj,Wan:2021vny,Wan:2023epq,Wan:2022uie,Wan:2019ake,Wan:2025zau,Wan:2025bdr,Wang:2020cme,Wan:2025ikc,Agaev:2022pis,Agaev:2025llz,Barsbay:2025vjq}. By constructing interpolating currents with the same quantum numbers as the tetraquark states, one can analyze the corresponding two-point correlation functions. Through the operator product expansion (OPE), nonperturbative effects are systematically incorporated via vacuum condensates. Applying a Borel transformation and quark--hadron duality allows the extraction of physical observables such as masses and pole residues. 
\cite{Chen:2010ze} 
\cite{Wang:2021qus}
\cite{Wang:2023jaw}

The QCD sum rule method has been widely applied to investigate the properties of hidden-charm tetraquark states. The pioneering study by Chen \textit{et al.}~\cite{Chen:2010ze} systematically constructed interpolating currents and analyzed the mass spectra for various $J^{PC}$ channels, laying a solid foundation for the field. More recently, investigations have extended to the vector $1^{--}$ sector~\cite{Wang:2021qus} and hidden-charm-hidden-strange systems~\cite{Wang:2023jaw}. Building upon these important works, we aim to further refine the theoretical predictions. Although Ref.~\cite{Chen:2010ze} provided a comprehensive analysis, the contribution of the dimension-6 three-gluon condensate, $\langle g_s^3 G^3 \rangle$, was not explicitly considered in the operator product expansion (OPE) at that stage. Incorporating this term, along with the updated non-perturbative vacuum condensates available today, allows for a more precise fine-tuning of the mass spectrum and a better analysis of the OPE convergence. Furthermore, distinct from recent studies focusing on the vector $1^{--}$ sector~\cite{Wang:2021qus} or the hidden-charm-hidden-strange systems~\cite{Wang:2023jaw}, the non-strange $1^{-+}$ tetraquarks ($cq\bar{c}\bar{q}$) possess unique exotic quantum numbers. A dedicated examination of their decay properties, constrained by the updated selection rules, is timely and necessary.

In this work, we focus on compact diquark–antidiquark configu-
rations, exploring various spin and color structures that may underlie the $1^{-+}$ hidden-charm
and hidden-bottom tetraquark states. within the QCD sum rule framework. We improve the precision of the previous theoretical calculations by explicitly extending the OPE series to include the three-gluon condensate contribution and by utilizing updated standard values for the non-perturbative vacuum condensates. These refinements allow for a more current and precise benchmark of the mass spectrum. Beyond spectroscopy, we also perform an analysis of the possible strong and electromagnetic decay modes, which may be accessible in future experiments at BESIII, Belle~II, LHCb, and future STCF.

This paper is organized as follows. In Sec.~\ref{Formalism}, we introduce the interpolating currents and the general formalism of QCD sum rules. Sec.~\ref{Numerical} presents the numerical analysis and discusses the mass spectra and pole residues. Sec.~\ref{Decay} addresses possible decay modes of the tetraquark states, and Sec.~\ref{Summary} contains a brief summary and concluding remarks.

\section{Formalism}\label{Formalism}

For the $1^{-+}$ hidden-charm tetraquark state, the corresponding interpolating currents can be constructed in the diquark–antidiquark configuration as follows:
\begin{eqnarray}\label{current_1-+}
j_\mu^A(x)&=&\frac{\epsilon_{a b c}\epsilon_{d e c}}{\sqrt{2}} {\Big\{}[q^T_a(x) C \gamma_5 c_b(x)][\bar{q}_d (x)\gamma_\mu\gamma_5 C \bar{c}^{T}_e(x)] \nonumber\\
&+&[q^T_a(x) C \gamma_\mu\gamma_5 c_b(x)][\bar{q}_d (x) \gamma_5 C \bar{c}^{ T}_e(x)] {\Big\}}\;,\label{Ja}\\
j_\mu^B(x)&=&\frac{\epsilon_{a b c}\epsilon_{d e c}}{\sqrt{2}} {\Big\{}[q^T_a(x) C \gamma_\nu\gamma_5 c_b(x)][\bar{q}_d (x)\sigma_{\mu\nu}\gamma_5 C \bar{c}^{T}_e(x)] \nonumber\\
&+&[q^T_a(x) C \sigma_{\mu\nu}\gamma_5 c_b(x)][\bar{q}_d (x) \gamma_\nu\gamma_5 C \bar{c}^{ T}_e(x)] {\Big\}}\;,\label{Jb}\\
j_\mu^C(x)&=&\frac{\epsilon_{a b c}\epsilon_{d e c}}{\sqrt{2}} {\Big\{}[q^T_a(x) C \gamma_\mu c_b(x)][\bar{q}_d (x) C \bar{c}^{T}_e(x)] \nonumber\\
&+&[q^T_a(x) C c_b(x)][\bar{q}_d (x) \gamma_\mu C \bar{c}^{ T}_e(x)] {\Big\}}\;,\label{Jc}\\
j_\mu^D(x)&=&\frac{\epsilon_{a b c}\epsilon_{d e c}}{\sqrt{2}} {\Big\{}[q^T_a(x) C \gamma_\nu c_b(x)][\bar{q}_d (x)\sigma_{\mu\nu} C \bar{c}^{T}_e(x)] \nonumber\\
&+&[q^T_a(x) C \sigma_{\mu\nu} c_b(x)][\bar{q}_d (x) \gamma_\nu C \bar{c}^{ T}_e(x)] {\Big\}}\;.\label{Jd}
\end{eqnarray}

In the above expressions, the subscripts $a,\cdots,e$ denote color indices, $q$ represents a light quark ($u$ or $d$), and $C$ denotes the charge-conjugation matrix.

Employing the interpolating currents in Eqs.~(\ref{Ja})--(\ref{Jd}), the two-point correlation functions are defined by
\begin{eqnarray}
\Pi_{\mu\nu}^k(q^2) &=& i \int d^4 x e^{i q \cdot x} \langle 0 | T \{ j_\mu^k(x),\;  j_\nu^k (0)^\dagger \} |0 \rangle \;,
\end{eqnarray}
where $ |0 \rangle$ denotes the physical vacuum, and the index $k$ runs from $A$ to $D$ Lorentz covariance dictates that the correlator admits the general decomposition
\begin{eqnarray}
\Pi_{\mu\nu}(q^2) &=&-\Big( g_{\mu \nu} - \frac{q_\mu q_\nu}{q^2}\Big) \Pi_1(q^2)+ \frac{q_\mu q_\nu}{q^2}\Pi_0(q^2)\;,
\end{eqnarray}
where $\Pi_1(q^2)$ and $\Pi_0(q^2)$ are invariant functions associated with intermediate hadronic states of the spin 1 and 0 mesons, respectively. Considering that we are focusing on the $1^{-+}$ meson, all the following calculations will be for $\Pi_1$, and we will omit the subscript $1$.

From the OPE side, the correlation function $\Pi(q^2)$ admits the dispersion representation
 \begin{eqnarray}
\Pi^{OPE}_k (q^2) = \int_{s_{min}}^{\infty} d s\frac{\rho^{OPE}_k(s)}{s - q^2}\; .
\label{OPE-hadron}
\end{eqnarray}
In the above expression, $s_{\rm min}$ denotes the kinematic threshold, which generally corresponds to the square of the sum of the current quark masses of the hadron~\cite{Albuquerque:2013ija}. The quantity $\rho^{\rm OPE}_k(s) = \mathrm{Im}\,\Pi^{\rm OPE}_k(s) / \pi$ is the spectral density obtained from the OPE, and
\begin{eqnarray}
\rho^{OPE}(s) & = & \rho^{pert}(s)  + \rho^{\langle \bar{q} q \rangle}(s)+\rho^{\langle G^2 \rangle}(s) + \rho^{\langle \bar{q} G q \rangle}(s)\nonumber\\
&+& \rho^{\langle \bar{q} q \rangle^2}(s)
+\rho^{\langle G^3 \rangle}(s)+\rho^{\langle \bar{q} q \rangle\langle \bar{q} G q \rangle}(s)\;. \label{rho-OPE}
\end{eqnarray}
The explicit analytical expressions of $\rho^{\rm OPE}_k(s)$ are derived and collected in Appendix~\ref{ana_exp}.

On the phenomenological side, following the standard QCD sum rule procedure and after isolating the ground-state tetraquark contribution, the correlation function $\Pi(q^2)$ can be expressed in terms of a dispersion integral over the physical region:
 \begin{eqnarray}
\Pi^{phen}_{k}(q^2) & = & \frac{\lambda_{k}^2}{M_k^2 - q^2} + \frac{1}{\pi} \int_{s_0}^\infty d s \frac{\rho_k(s)}{s - q^2} \; . \label{hadron}
\end{eqnarray}
Here, $M$ represents the mass of the tetraquark state, $\lambda$ is the pole residue associated with the interpolating current, and $\rho(s)$ denotes the spectral density, which includes contributions from higher excited states as well as the continuum above the threshold $s_0$.

After performing a Borel transform on Eqs.~(\ref{OPE-hadron}) and (\ref{hadron}) and matching the operator product expansion with the phenomenological side of the correlation function $\Pi(q^2)$, the mass of the tetraquark state is obtained as
\begin{eqnarray}
M_k(s_0, M_B^2) = \sqrt{- \frac{L^k_{1}(s_0, M_B^2)}{L^k_{0}(s_0, M_B^2)}} \; . \label{mass-Eq}
\end{eqnarray}
Here $L_0$ and $L_1$ are respectively defined as
\begin{eqnarray}
L^k_{,\;0}(s_0, M_B^2) =  \int_{s_{min}}^{s_0} d s \; \rho^{OPEk}_{k}(s) e^{-
s / M_B^2}   \;,  \label{L0}
\end{eqnarray}
and
\begin{eqnarray}
L^k_{1}(s_0, M_B^2) =
\frac{\partial}{\partial{\frac{1}{M_B^2}}}{L^k_{0}(s_0, M_B^2)} \; .
\end{eqnarray}

\section{Numerical analysis}\label{Numerical}

In performing the numerical analysis, we utilize the widely accepted input parameters as commonly adopted in QCD sum rule studies \cite{Chung:1981wm,Albuquerque:2013ija,Wang:2013vex,Reinders:1984sr,P.Col,Narison:1989aq}:
\begin{eqnarray}
\begin{aligned}
&m_c(m_c)=\overline{m}_c=(1.275\pm0.025)\; \text{GeV}\;,&&m_b(m_b)=\overline{m}_b=(4.18\pm0.03)\; \text{GeV}   \; ,\\
& \langle \bar{q} q \rangle = - (0.24 \pm 0.01)^3 \; \text{GeV}^3 \; ,& & \langle g_s^2 G^2 \rangle = (0.88\pm0.25) \; \text{GeV}^4 \; , \\
& \langle \bar{q} g_s \sigma \cdot G q \rangle = m_0^2 \langle\bar{q} q \rangle; , & & \langle g_s^3 G^3 \rangle = (0.045 \pm 0.013) \;\text{GeV}^6 \; ,\\
& m_0^2 = (0.8 \pm 0.1) \; \text{GeV}^2\; .
\end{aligned}
\end{eqnarray}
In the above, $\overline{m}_c$ and $\overline{m}_b$ correspond to the heavy-quark running masses in the $\overline{\rm MS}$ scheme, and the light quark masses are taken in the chiral limit, $m_q = 0$.

In addition, the two parameters $s_0$ and $M_B^2$, which are introduced in the construction of the sum rules, need to be determined. They can be fixed following the standard procedures by satisfying the following two criteria~\cite{Shifman,Reinders:1984sr,P.Col}. The first criterion ensures the convergence of the OPE, which is assessed by comparing the relative contributions of higher-dimensional condensates to the total OPE contribution. A reliable Borel window for $M_B^2$ is then chosen to maintain this convergence. The second criterion requires that the pole contribution (PC) accounts for more than 50\% of the total contribution for tetraquark states~\cite{Shifman,Reinders:1984sr,P.Col}. These two criteria can be expressed mathematically as:
\begin{eqnarray}
  R^{OPE}= \left| \frac{L_{0}^{k\;,dim=8}(s_0, M_B^2)}{L_{0}^k(s_0, M_B^2)}\right|\, ,
\end{eqnarray}
\begin{eqnarray}
  R^{PC} = \frac{L_{0}^k(s_0, M_B^2)}{L_{0}^k(\infty, M_B^2)} \; . \label{RatioPC}
\end{eqnarray}

To determine a suitable value for the continuum threshold $s_0$, we follow a procedure similar to that employed in Refs.~\cite{Qiao:2013dda,Tang:2016pcf,Qiao:2013raa}. Specifically, one seeks the value of $s_0$ that provides an optimal Borel window for the mass curve of the tetraquark state. Within this window, the physical quantity of interest—the tetraquark mass—should exhibit minimal dependence on the Borel parameter $M_B^2$. In practice, we vary $\sqrt{s_0}$ by $\pm 0.1$~GeV to establish the corresponding lower and upper bounds, thereby estimating the uncertainties associated with $\sqrt{s_0}$~\cite{Wan:2020oxt,Wan:2020fsk}.

\begin{figure}
\includegraphics[width=6.8cm]{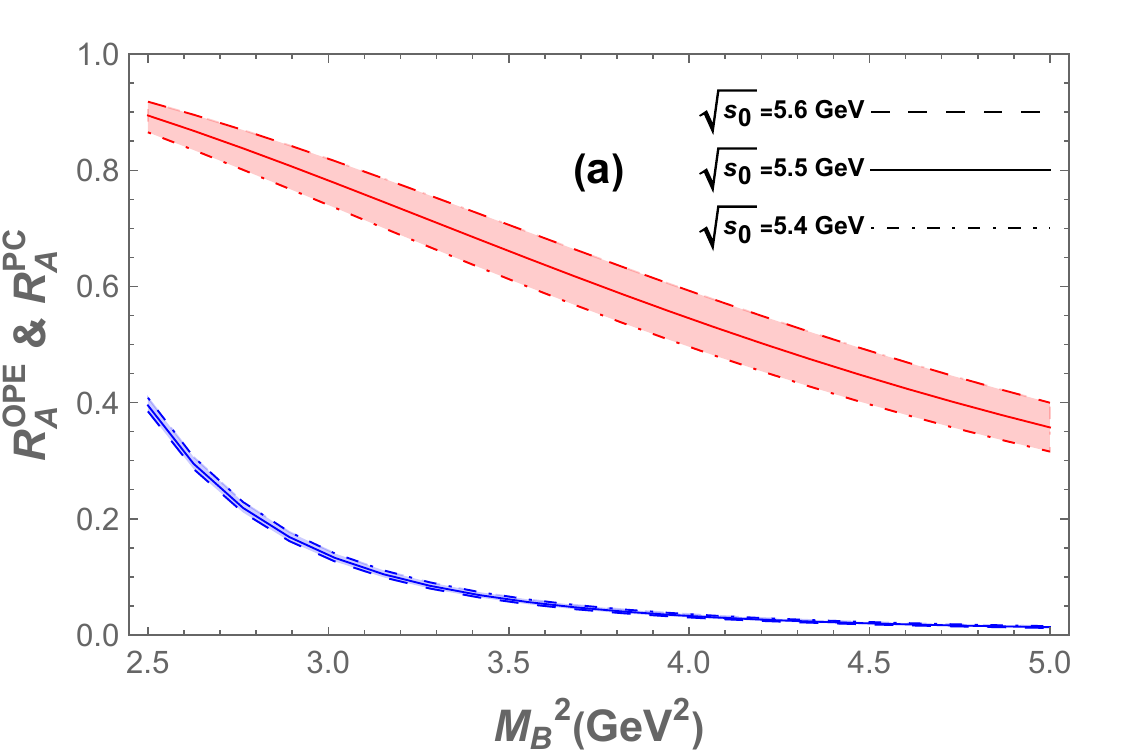}
\includegraphics[width=6.8cm]{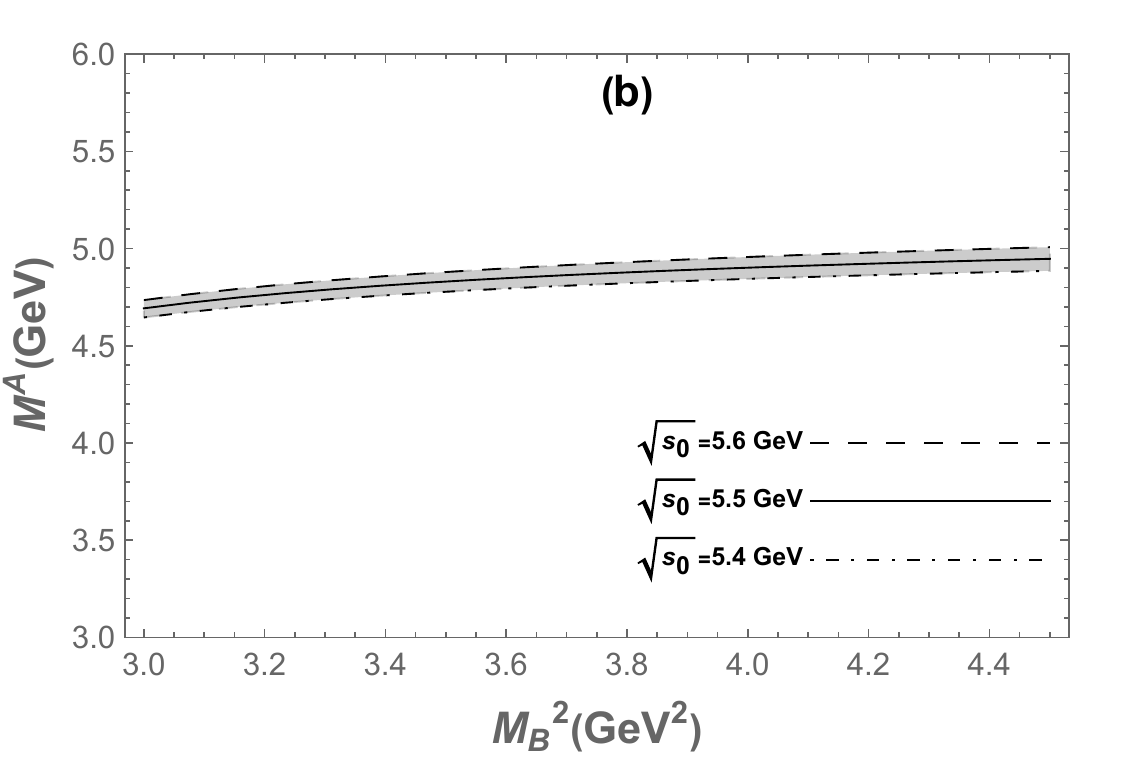}
\caption{ (a) The ratios of ${R_{A}^{OPE}}$ and ${R_{A}^{PC}}$ as functions of the Borel parameter $M_B^2$ for different values of $\sqrt{s_0}$, where blue lines represent ${R_{A}^{OPE}}$ and red lines denote ${R_{A}^{PC}}$ . (b) The mass $M^{A}$ as a function of the Borel parameter $M_B^2$ for different values of $\sqrt{s_0}$.} \label{figA}
\end{figure}

With the above preparations, the mass spectrum of the $1^{-+}$ hidden-charm tetraquark states can be numerically evaluated. As an example, for the interpolating current in Eq.~(\ref{Ja}), the ratios $R^{\rm OPE,A}$ and $R^{\rm PC,A}$ are shown as functions of the Borel parameter $M_B^2$ in Fig.~\ref{figA}(a) for different values of $\sqrt{s_0}$, namely $5.4$, $5.5$, and $5.6$~GeV. The corresponding dependence of the mass $M^{A}$ on $M_B^2$ is displayed in Fig.~\ref{figA}(b). The optimal Borel window is determined to be $3.2 \le M_B^2 \le 4.2~\text{GeV}^2$, within which the mass of the tetraquark state is extracted as
\begin{eqnarray}
M^{A} &=& (4.83 \pm 0.15)~\text{GeV}\,.
\label{m1}
\end{eqnarray}

Following the same analysis, and with the OPE, pole contributions, and the masses as functions of the Borel parameter $M_B^2$ provided in Appendix~\ref{pictures}, the masses of the $1^{-+}$ hidden-charm tetraquark states corresponding to the currents in Eqs.~(\ref{Jb})--(\ref{Jd}) can be extracted and are summarized in Table~\ref{mass}.
Similarly, the $1^{-+}$ hidden-bottom tetraquark states can be analyzed. By employing the same analytical expressions as in the hidden-charm case, with $m_c$ replaced by $m_b$, the corresponding masses are obtained and also summarized in Table~\ref{mass}.
The errors in the present analysis are mainly due to the uncertainties associated with the quark masses, the vacuum condensates, and the continuum threshold parameter $\sqrt{s_0}$.

\begin{table}
\begin{center}
\renewcommand\tabcolsep{10pt}
\caption{The continuum thresholds, Borel parameters, and predicted masses of hidden-charm baryonium.}\label{mass}
\begin{tabular}{ccccc}\hline\hline
                    &Current   & $\sqrt{s_0}\;(\text{GeV})$     &$M_B^2\;(\text{GeV}^2)$ &$M^X\;(\text{GeV})$       \\ \hline
$c$-sector    &$A$        & $5.5\pm0.1$                             &$3.2-4.2$                      &$4.83\pm0.15$         \\
                     &$B$        & $5.6\pm0.1$                             &$3.0-4.6$                      &$4.88\pm0.18$          \\
                     &$C$        & $5.4\pm0.1$                             &$2.9-4.1$                      &$4.72\pm0.16$ \\
                    &$D$        & $5.4\pm0.1$                           &$3.1-3.9$                     &$4.79\pm0.12$           \\ \hline
$b$-sector     &$A$       & $12.0\pm0.1$                         &$9.5-12.0$                   &$11.08\pm0.16$           \\
                     &$B$        & $12.1\pm0.1$                         &$10.0-12.4$                  &$11.16\pm0.14$          \\      
                     &$C$        & $11.9\pm0.1$                          &$9.3-11.8$                      &$10.99\pm0.16$          \\  
                     &$D$        & $11.9\pm0.1$                          &$9.4-11.6$                      &$11.03\pm0.15$          \\                                                           
\hline
 \hline
\end{tabular}
\end{center}
\end{table}

\section{Decay Modes of $1^{-+}$ Hidden-Charm Tetraquark States}\label{Decay}

After determining the mass spectrum of the $1^{-+}$ hidden-charm tetraquark states, it is instructive to discuss their possible decay modes. Given the quark content $c\bar{c}q\bar{q}$ ($q=u,d$) and the exotic quantum numbers $J^{PC}=1^{-+}$, the decay patterns are strongly constrained by both conservation laws and phase space considerations.

\subsection{Strong Decays}

The dominant decay channels are expected to be governed by the strong interaction. For tetraquark states with masses around 4.7--4.9~GeV, the kinematically allowed two-body strong decays include:
\begin{itemize}
    \item $1^{-+} \to \eta_c \pi$, where the tetraquark decays into a pseudoscalar charmonium and a light pseudoscalar meson;
    \item $1^{-+} \to J/\psi \rho$, a decay into a vector charmonium and a light vector meson;
    \item $1^{-+} \to D \bar{D}^{*}$ or $D^* \bar{D}^*$, involving open-charm meson pairs.
\end{itemize}

These decay modes are favored since they respect conservation of angular momentum, parity, and charge conjugation. Three-body decays such as $1^{-+} \to J/\psi \pi \pi$ or $\eta_c \pi \pi$ are also possible but are typically suppressed relative to two-body channels due to reduced phase space and additional coupling constants.

\subsection{Electromagnetic and Weak Decays}

Electromagnetic decays, e.g., $1^{-+} \to J/\psi \gamma$, are expected to have smaller branching ratios compared with strong decays but can provide clean experimental signatures. Weak decays are negligible due to the much longer time scale associated with $c$-quark decay.

\subsection{Experimental Prospects}

The decay channels involving $J/\psi$ in the final state are particularly promising for experimental detection, as $J/\psi$ can be efficiently reconstructed via its dilepton decays. Thus, experiments at BESIII, BelleII, and LHCb have the potential to search for these exotic states through the proposed decay channels. The identification of specific $1^{-+}$ candidates can be aided by analyzing invariant mass distributions and angular correlations among the decay products.

In summary, the $1^{-+}$ hidden-charm tetraquark states are expected to predominantly decay via strong two-body channels into charmonium plus light mesons or open-charm meson pairs, with electromagnetic channels providing complementary but cleaner signatures for experimental searches.

\section{Results and Discussion}\label{Summary}

Using QCD sum rules, we have evaluated the masses of the $1^{-+}$ hidden-charm and hidden-bottom tetraquark states. The results for the hidden-charm sector are summarized in Table~\ref{mass}, indicating the possible existence of four states with masses $(4.83\pm0.15)$, $(4.88\pm0.18)$, $(4.72\pm0.16)$, and $(4.79\pm0.12)$ GeV, respectively.
The corresponding hidden-bottom partners, obtained by replacing $m_c$ with $m_b$ in the analytical expressions, are found to be $(11.08\pm0.16)$, $(11.16\pm0.14)$, $(10.99\pm0.16)$, and $(11.03\pm0.15)$ GeV, respectively. The quoted uncertainties mainly originate from the variations in the heavy-quark masses, the values of vacuum condensates, the Borel parameter, and the continuum threshold $\sqrt{s_0}$. The relatively stable behavior of the extracted masses within the chosen Borel windows ensures the reliability of our results.

Furthermore, we have analyzed the possible strong and electromagnetic decay modes of the $1^{-+}$ tetraquark states. In the hidden-charm sector, these states can decay into final states containing charmonium and light mesons, such as $\eta_c\,\pi$, $J/\psi\,\rho$, as well as open-charm meson pairs like $D\bar{D}^*$ and $D^*\bar{D}^*$. In addition, electromagnetic transitions to lower charmonium states accompanied by a photon are also possible. In particular, decay chains involving
\begin{equation*}
J/\psi \to \ell^+\ell^-, \quad \ell = e,\, \mu,
\end{equation*}
provide clean experimental signatures, which can be efficiently reconstructed in current and future experiments.

The mass range and decay patterns obtained in this work provide valuable theoretical guidance for the experimental exploration of exotic hadrons with quantum numbers $J^{PC}=1^{-+}$. These results are expected to be tested at facilities such as BESIII, Belle~II and LHCb, as well as at upcoming high-luminosity and high-energy experiments and future STCF \cite{Achasov:2023gey}. The confirmation of such states would represent a significant step forward in the understanding of multiquark dynamics and the nonperturbative structure of QCD.

We hope that the present study will stimulate further experimental and theoretical investigations of exotic hadron spectroscopy and contribute to clarifying the internal structure of tetraquark states with exotic quantum numbers.

\vspace{.5cm} {\bf Acknowledgments} \vspace{.5cm}

This work was supported in part by the National Natural Science Foundation of China under Grants 12575106 and 12147214, and Specific Fund of Fundamental Scientific Research Operating Expenses for Undergraduate Universities in Liaoning Province under Grants No. LJ212410165019.


\begin{widetext}
\appendix

\section{The spectral densities for $1^{-+}$ hidden-charm tetraquark states}\label{ana_exp}
\subsection{The spectral densities for current in Eq.~(\ref{Ja})}

\begin{eqnarray}
\rho^{pert}(s)&=&\int_{\alpha_{min}}^{\alpha_{max}} d \alpha \int_{\beta_{min}}^{1-\alpha} d \beta \bigg{\{}-\frac{{\cal F}_{\alpha \beta}^{3}(\alpha+\beta-1)}{3072\pi^{6} \alpha^{3} \beta^{3}}\nonumber\\
&\times&\bigg(3{\cal F}_{\alpha \beta}(\alpha+\beta+1)+2m_Q^2(\alpha+\beta-1)^2\bigg)\bigg{\}}\; ,\\
\rho^{\langle \bar{q} q\rangle}(s)&=&\int_{\alpha_{min}}^{\alpha_{max}} d \alpha \int_{\beta_{min}}^{1-\alpha} d \beta \bigg{\{}
-\frac{{\cal F}_{\alpha \beta}^{2}m_Q \langle \bar{q} q\rangle(\alpha+\beta-1)(\alpha+\beta)}{32\pi^4\alpha^2\beta^2 }\bigg{\}}\; ,\\
\rho^{\langle G^2 \rangle}(s)&=&\frac{\langle G^2 \rangle}{\pi^{6}}\int_{\alpha_{min}}^{\alpha_{max}} d \alpha \int_{\beta_{min}}^{1-\alpha} d \beta \bigg{\{}\frac{{\cal F}_{\alpha \beta} }{73728 \alpha^{2} \beta^{2}}\bigg((\alpha+\beta-1)^{2} m_Q^{2}(17\alpha \nonumber \\
&+&17\beta-5)-12{\cal F}_{\alpha \beta} (2\alpha^2+\beta(2\beta-1)+\alpha(4\beta-1)) \bigg) \nonumber\\
&+&\frac{(\alpha+\beta-1)}{18432 \alpha^{3} \beta^{3}}\bigg( m_Q^{4}(\alpha+\beta-1)^2 (\alpha^3+\beta^3)+3{\cal F}_{\alpha \beta} m_Q^2(3\alpha^4 \nonumber\\
&+&4\alpha^3\beta+\beta^2+3\beta^4+2\alpha\beta^2(2\beta-1) +\alpha^2(1-2\beta+2\beta^2) )\bigg)\bigg{\}}\; ,\\
\rho^{\langle \bar{q} G q \rangle} (s)&=&\frac{\langle \bar{q} G q \rangle}{64 \pi^{6}}\int_{\alpha_{min}}^{\alpha_{max}} d \alpha \int_{\beta_{min}}^{1-\alpha} d \beta \bigg{\{}  \frac{m_Q {\cal F}_{\alpha \beta}(\alpha+\beta)}{\alpha\beta}  \bigg{\}}\;,\\
\rho^{\langle \bar{q} q\rangle^2}(s)&=&\frac{\langle \bar{q} q\rangle^2}{24\pi^2} \int_{\alpha_{min}}^{\alpha_{max}} d \alpha \bigg{\{}{\cal H}_\alpha -2m_Q^2 \bigg{\}} \;,\\
\rho^{\langle G^3 \rangle}(s) &=&\frac{-\langle G^3 \rangle}{12288\pi^{6}}\int_{\alpha_{min}}^{\alpha_{max}} \frac{d \alpha}{ \alpha^3} \int_{\beta_{min}}^{1-\alpha} \frac{d \beta}{ \beta^3} (\alpha+\beta-1) \bigg{\{}  {\cal F}_{\alpha \beta}(\alpha+\beta) (\alpha+\beta+1) (\alpha^2\nonumber\\
&-&\alpha\beta+\beta^2) + m_Q^2(3\alpha^5+\alpha^3(\beta-1)^2 +4\alpha^4\beta+\beta^3 \nonumber\\
&+&\alpha^2\beta^3 +3\beta^5+2\alpha\beta^3(2\beta-1) )   \bigg{\}}\; ,\\
\rho^{\langle \bar{q} q \rangle\langle \bar{q} G q \rangle}(s)&=&-\int_{\alpha_{min}}^{\alpha_{max}} d \alpha \frac{\langle \bar{q} q \rangle\langle \bar{q} G q \rangle(\alpha-1)\alpha}{24\pi^2}\;,
\end{eqnarray}
\begin{eqnarray}
\Pi^{\langle \bar{q} q \rangle\langle \bar{q} G q \rangle}(M_B^2)&=&\frac{m_Q^2\langle \bar{q} q \rangle\langle \bar{q} G q \rangle\;Exp(\frac{m_Q^2}{\alpha(\alpha-1)M_B^2})}{48\pi^2M_B^2}\int_{0}^{1} d \alpha \frac{3\alpha(\alpha-1)M_B^2-2m_Q^2}{\alpha(\alpha-1)}\;,
\end{eqnarray}
where $M_B$ is the Borel parameter introduced through the Borel transformation, and $Q=c$ or $b$ denotes the heavy-quark flavor under consideration. The quantity $\Pi(M_B^2)$ represents the contribution to the correlation function that has no imaginary part but acquires a non-vanishing magnitude after the Borel transformation. In the following, we further introduce the definitions:
\begin{eqnarray}
{\cal F}_{\alpha \beta} &=& (\alpha + \beta) m_Q^2 - \alpha \beta s \; , {\cal H}_\alpha  = m_Q^2 - \alpha (1 - \alpha) s \; , \\
\alpha_{min} &=& \left(1 - \sqrt{1 - 4 m_Q^2/s} \right) / 2, \; , \alpha_{max} = \left(1 + \sqrt{1 - 4 m_Q^2 / s} \right) / 2  \; , \\
\beta_{min} &=& \alpha m_Q^2 /(s \alpha - m_Q^2).
\end{eqnarray}

\subsection{The spectral densities for current in Eq.~(\ref{Jb})}

\begin{eqnarray}
\rho^{pert}(s)&=&\int_{\alpha_{min}}^{\alpha_{max}} d \alpha \int_{\beta_{min}}^{1-\alpha} d \beta \bigg{\{}-\frac{{\cal F}_{\alpha \beta}^{3}(\alpha+\beta-1)}{1024\pi^{6} \alpha^{3} \beta^{3}}\nonumber\\
&\times&\bigg(3{\cal F}_{\alpha \beta}(\alpha+\beta+1)+2m_Q^2(\alpha+\beta-1)^2\bigg)\bigg{\}}\; ,\\
\rho^{\langle \bar{q} q\rangle}(s)&=&\int_{\alpha_{min}}^{\alpha_{max}} d \alpha \int_{\beta_{min}}^{1-\alpha} d \beta \bigg{\{}
\frac{3{\cal F}_{\alpha \beta}^{2}m_Q \langle \bar{q} q\rangle(\alpha+\beta-1)(\alpha+\beta)}{32\pi^4\alpha^2\beta^2 }\bigg{\}}\; ,\\
\rho^{\langle G^2 \rangle}(s)&=&\frac{\langle G^2 \rangle}{\pi^{6}}\int_{\alpha_{min}}^{\alpha_{max}} d \alpha \bigg{\{} \int_{\beta_{min}}^{1-\alpha} d \beta \bigg( \frac{{\cal F}_{\alpha \beta} }{73728 \alpha^{2} \beta^{2}}\Big(- m_Q^{2}(13\alpha^3 \nonumber \\
&+&(13\beta+29)(\beta-1)^2 +\alpha^2 (51\beta+3) + \alpha(51\beta^2-54\beta-45) ) \nonumber\\
&-&12{\cal F}_{\alpha \beta} (\alpha^2+\beta(\beta-1)+\alpha(4\beta-1)) \Big) \nonumber\\
&-&\frac{(\alpha+\beta-1)}{6144 \alpha^{3} \beta^{3}}\Big( m_Q^{4}(\alpha+\beta-1)^2 (\alpha^3+\beta^3)+3{\cal F}_{\alpha \beta} m_Q^2(3\alpha^4 \nonumber\\
&+&4\alpha^3\beta+\beta^2+3\beta^4+2\alpha\beta^2(2\beta-1) +\alpha^2(1-2\beta+2\beta^2) )\Big) \bigg)\nonumber\\
& &-\frac{{\cal H}_\alpha^2}{6144\alpha(\alpha-1)}  \bigg{\}}\; ,\\
\rho^{\langle \bar{q} G q \rangle} (s)&=&-\frac{3\langle \bar{q} G q \rangle}{64 \pi^{6}}\int_{\alpha_{min}}^{\alpha_{max}} d \alpha \int_{\beta_{min}}^{1-\alpha} d \beta \bigg{\{}  \frac{m_Q {\cal F}_{\alpha \beta}(\alpha+\beta)}{\alpha\beta} \bigg{\}}\;,\\
\rho^{\langle \bar{q} q\rangle^2}(s)&=&\frac{\langle \bar{q} q\rangle^2}{8\pi^2} \int_{\alpha_{min}}^{\alpha_{max}} d \alpha \bigg{\{}{\cal H}_\alpha -2m_Q^2 \bigg{\}} \;,\\
\rho^{\langle G^3 \rangle}(s) &=&\frac{-\langle G^3 \rangle}{4096\pi^{6}}\int_{\alpha_{min}}^{\alpha_{max}} \frac{d \alpha}{ \alpha^3} \int_{\beta_{min}}^{1-\alpha} \frac{d \beta}{ \beta^3} (\alpha+\beta-1) \bigg{\{}  {\cal F}_{\alpha \beta}(\alpha+\beta) (\alpha+\beta+1) (\alpha^2\nonumber\\
&-&\alpha\beta+\beta^2) + m_Q^2(3\alpha^5+\alpha^3(\beta-1)^2 +4\alpha^4\beta+\beta^3 \nonumber\\
&+&\alpha^2\beta^3 +3\beta^5+2\alpha\beta^3(2\beta-1)  )  \bigg{\}}\; ,\\
\rho^{\langle \bar{q} q \rangle\langle \bar{q} G q \rangle}(s)&=&-\int_{\alpha_{min}}^{\alpha_{max}} d \alpha \frac{\langle \bar{q} q \rangle\langle \bar{q} G q \rangle(\alpha-1)\alpha}{8\pi^2}\;,
\end{eqnarray}
\begin{eqnarray}
\Pi^{\langle \bar{q} q \rangle\langle \bar{q} G q \rangle}(M_B^2)&=&\frac{m_Q^2\langle \bar{q} q \rangle\langle \bar{q} G q \rangle\;Exp(\frac{m_Q^2}{\alpha(\alpha-1)M_B^2})}{16\pi^2M_B^2}\int_{0}^{1} d \alpha \frac{3\alpha(\alpha-1)M_B^2-2m_Q^2}{\alpha(\alpha-1)}\;.
\end{eqnarray}

\subsection{The spectral densities for current in Eq.~(\ref{Jc})}

\begin{eqnarray}
\rho^{pert}(s)&=&\int_{\alpha_{min}}^{\alpha_{max}} d \alpha \int_{\beta_{min}}^{1-\alpha} d \beta \bigg{\{}\frac{{\cal F}_{\alpha \beta}^{3}(\alpha+\beta-1)}{3072\pi^{6} \alpha^{3} \beta^{3}}\nonumber\\
&\times&\bigg(3{\cal F}_{\alpha \beta}(\alpha+\beta+1)+2m_Q^2(\alpha+\beta-1)^2\bigg)\bigg{\}}\; ,\\
\rho^{\langle \bar{q} q\rangle}(s)&=&\int_{\alpha_{min}}^{\alpha_{max}} d \alpha \int_{\beta_{min}}^{1-\alpha} d \beta \bigg{\{}
-\frac{{\cal F}_{\alpha \beta}^{2}m_Q \langle \bar{q} q\rangle(\alpha+\beta-1)(\alpha+\beta)}{32\pi^4\alpha^2\beta^2 }\bigg{\}}\; ,\\
\rho^{\langle G^2 \rangle}(s)&=&\frac{\langle G^2 \rangle}{\pi^{6}}\int_{\alpha_{min}}^{\alpha_{max}} d \alpha \int_{\beta_{min}}^{1-\alpha} d \beta \bigg{\{}-\frac{{\cal F}_{\alpha \beta} }{73728 \alpha^{2} \beta^{2}}\bigg((\alpha+\beta-1)^{2} m_Q^{2}(17\alpha \nonumber \\
&+&17\beta-5)-12{\cal F}_{\alpha \beta} (2\alpha^2+\beta(2\beta-1)+\alpha(4\beta-1)) \bigg) \nonumber\\
&+&\frac{(\alpha+\beta-1)}{18432 \alpha^{2} \beta^{2}}\bigg( m_Q^{4}(\alpha+\beta-1)^2 (\alpha^3+\beta^3)+3{\cal F}_{\alpha \beta} m_Q^2(3\alpha^4 \nonumber\\
&+&4\alpha^3\beta+\beta^2+3\beta^4+2\alpha\beta^2(2\beta-1) +\alpha^2(1-2\beta+2\beta^2) )\bigg)\bigg{\}}\; ,\\
\rho^{\langle \bar{q} G q \rangle} (s)&=&-\frac{\langle \bar{q} G q \rangle}{64 \pi^{6}}\int_{\alpha_{min}}^{\alpha_{max}} d \alpha \int_{\beta_{min}}^{1-\alpha} d \beta \bigg{\{}  \frac{m_Q {\cal F}_{\alpha \beta}(\alpha+\beta)}{\alpha\beta}  \bigg{\}}\;,\\
\rho^{\langle \bar{q} q\rangle^2}(s)&=&\frac{\langle \bar{q} q\rangle^2}{24\pi^2} \int_{\alpha_{min}}^{\alpha_{max}} d \alpha \bigg{\{}{\cal H}_\alpha -2m_Q^2 \bigg{\}} \;,\\
\rho^{\langle G^3 \rangle}(s) &=&\frac{\langle G^3 \rangle}{12288\pi^{6}}\int_{\alpha_{min}}^{\alpha_{max}} \frac{d \alpha}{ \alpha^3} \int_{\beta_{min}}^{1-\alpha} \frac{d \beta}{ \beta^3} (\alpha+\beta-1) \bigg{\{}  {\cal F}_{\alpha \beta}(\alpha+\beta) (\alpha+\beta+1) (\alpha^2\nonumber\\
&-&\alpha\beta+\beta^2) + m_Q^2(3\alpha^5+\alpha^3(\beta-1)^2 +4\alpha^4\beta+\beta^3 \nonumber\\
&+&\alpha^2\beta^3 +3\beta^5+2\alpha\beta^3(2\beta-1) )   \bigg{\}}\; ,\\
\rho^{\langle \bar{q} q \rangle\langle \bar{q} G q \rangle}(s)&=&\int_{\alpha_{min}}^{\alpha_{max}} d \alpha \frac{\langle \bar{q} q \rangle\langle \bar{q} G q \rangle(\alpha-1)\alpha}{24\pi^2}\;,
\end{eqnarray}
\begin{eqnarray}
\Pi^{\langle \bar{q} q \rangle\langle \bar{q} G q \rangle}(M_B^2)&=&-\frac{m_Q^2\langle \bar{q} q \rangle\langle \bar{q} G q \rangle\;Exp(\frac{m_Q^2}{\alpha(\alpha-1)M_B^2})}{48\pi^2M_B^2}\int_{0}^{1} d \alpha \frac{3\alpha(\alpha-1)M_B^2-2m_Q^2}{\alpha(\alpha-1)}\;.
\end{eqnarray}

\subsection{The spectral densities for current in Eq.~(\ref{Jd})}

\begin{eqnarray}
\rho^{pert}(s)&=&\int_{\alpha_{min}}^{\alpha_{max}} d \alpha \int_{\beta_{min}}^{1-\alpha} d \beta \bigg{\{}\frac{{\cal F}_{\alpha \beta}^{3}(\alpha+\beta-1)}{1024\pi^{6} \alpha^{3} \beta^{3}}\nonumber\\
&\times&\bigg(3{\cal F}_{\alpha \beta}(\alpha+\beta+1)+2m_Q^2(\alpha+\beta-1)^2\bigg)\bigg{\}}\; ,\\
\rho^{\langle \bar{q} q\rangle}(s)&=&\int_{\alpha_{min}}^{\alpha_{max}} d \alpha \int_{\beta_{min}}^{1-\alpha} d \beta \bigg{\{}
\frac{3{\cal F}_{\alpha \beta}^{2}m_Q \langle \bar{q} q\rangle(\alpha+\beta-1)(\alpha+\beta)}{32\pi^4\alpha^2\beta^2 }\bigg{\}}\; ,\\
\rho^{\langle G^2 \rangle}(s)&=&\frac{\langle G^2 \rangle}{\pi^{6}}\int_{\alpha_{min}}^{\alpha_{max}} d \alpha  \int_{\beta_{min}}^{1-\alpha} d \beta \bigg( \frac{{\cal F}_{\alpha \beta} }{73728 \alpha^{2} \beta^{2}}\Big(m_Q^{2}(13\alpha^3 \nonumber \\
&+&(13\beta+29)(\beta-1)^2 +\alpha^2 (51\beta+3)  +\alpha(51\beta^2-54\beta-45) ) \nonumber\\
&+&12{\cal F}_{\alpha \beta} (\alpha^2+\beta(\beta-1)+\alpha(4\beta-1)) \Big) \nonumber\\
&-&\frac{(\alpha+\beta-1)}{6144 \alpha^{3} \beta^{3}}\Big( m_Q^{4}(\alpha+\beta-1)^2 (\alpha^3+\beta^3)+3{\cal F}_{\alpha \beta} m_Q^2(3\alpha^4 \nonumber\\
&+&4\alpha^3\beta+\beta^2+3\beta^4+2\alpha\beta^2(2\beta-1) +\alpha^2(1-2\beta+2\beta^2) )\Big) \bigg)\; ,\\
\rho^{\langle \bar{q} G q \rangle} (s)&=&-\frac{3\langle \bar{q} G q \rangle}{64 \pi^{6}}\int_{\alpha_{min}}^{\alpha_{max}} d \alpha \int_{\beta_{min}}^{1-\alpha} d \beta \bigg{\{}  \frac{m_Q {\cal F}_{\alpha \beta}(\alpha+\beta)}{\alpha\beta} \bigg{\}}\;,\\
\rho^{\langle \bar{q} q\rangle^2}(s)&=&-\frac{\langle \bar{q} q\rangle^2}{8\pi^2} \int_{\alpha_{min}}^{\alpha_{max}} d \alpha \bigg{\{}{\cal H}_\alpha -2m_Q^2 \bigg{\}} \;,\\
\rho^{\langle G^3 \rangle}(s) &=&\frac{\langle G^3 \rangle}{4096\pi^{6}}\int_{\alpha_{min}}^{\alpha_{max}} \frac{d \alpha}{ \alpha^3} \int_{\beta_{min}}^{1-\alpha} \frac{d \beta}{ \beta^3} (\alpha+\beta-1) \bigg{\{}  {\cal F}_{\alpha \beta}(\alpha+\beta) (\alpha+\beta+1) (\alpha^2\nonumber\\
&-&\alpha\beta+\beta^2) + m_Q^2(3\alpha^5+\alpha^3(\beta-1)^2 +4\alpha^4\beta+\beta^3 \nonumber\\
&+&\alpha^2\beta^3 +3\beta^5+2\alpha\beta^3(2\beta-1)  )  \bigg{\}}\; ,\\
\rho^{\langle \bar{q} q \rangle\langle \bar{q} G q \rangle}(s)&=&\int_{\alpha_{min}}^{\alpha_{max}} d \alpha \frac{\langle \bar{q} q \rangle\langle \bar{q} G q \rangle(\alpha-1)\alpha}{8\pi^2}\;,
\end{eqnarray}
\begin{eqnarray}
\Pi^{\langle \bar{q} q \rangle\langle \bar{q} G q \rangle}(M_B^2)&=-&\frac{m_Q^2\langle \bar{q} q \rangle\langle \bar{q} G q \rangle\;Exp(\frac{m_Q^2}{\alpha(\alpha-1)M_B^2})}{16\pi^2M_B^2}\int_{0}^{1} d \alpha \frac{3\alpha(\alpha-1)M_B^2-2m_Q^2}{\alpha(\alpha-1)}\;.
\end{eqnarray}

\section{pictures}\label{pictures}
For the $1^{-+}$ hidden-charm tetraquark states in Eqs.~(\ref{Jb})-(\ref{Jd}), the OPE, pole contribution and the masses as functions of Borel parameter $M_B^2$ are given in Figs.~\ref{figB}-\ref{figD}. For the $1^{-+}$ hidden-bottom tetraquark states in Eqs.~(\ref{Ja})-(\ref{Jd}), the OPE, pole contribution and the masses as functions of Borel parameter $M_B^2$ are given in Figs.~\ref{figAb}-\ref{figDb}.

\begin{figure}[h]
\includegraphics[width=6.8cm]{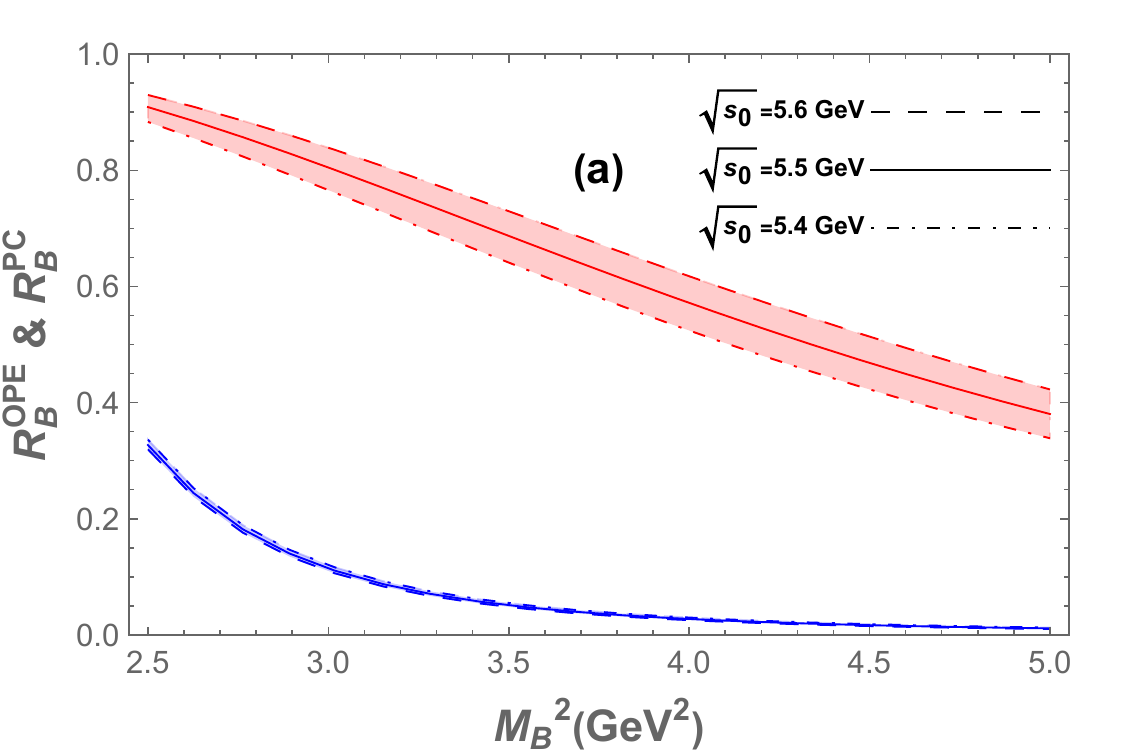}
\includegraphics[width=6.8cm]{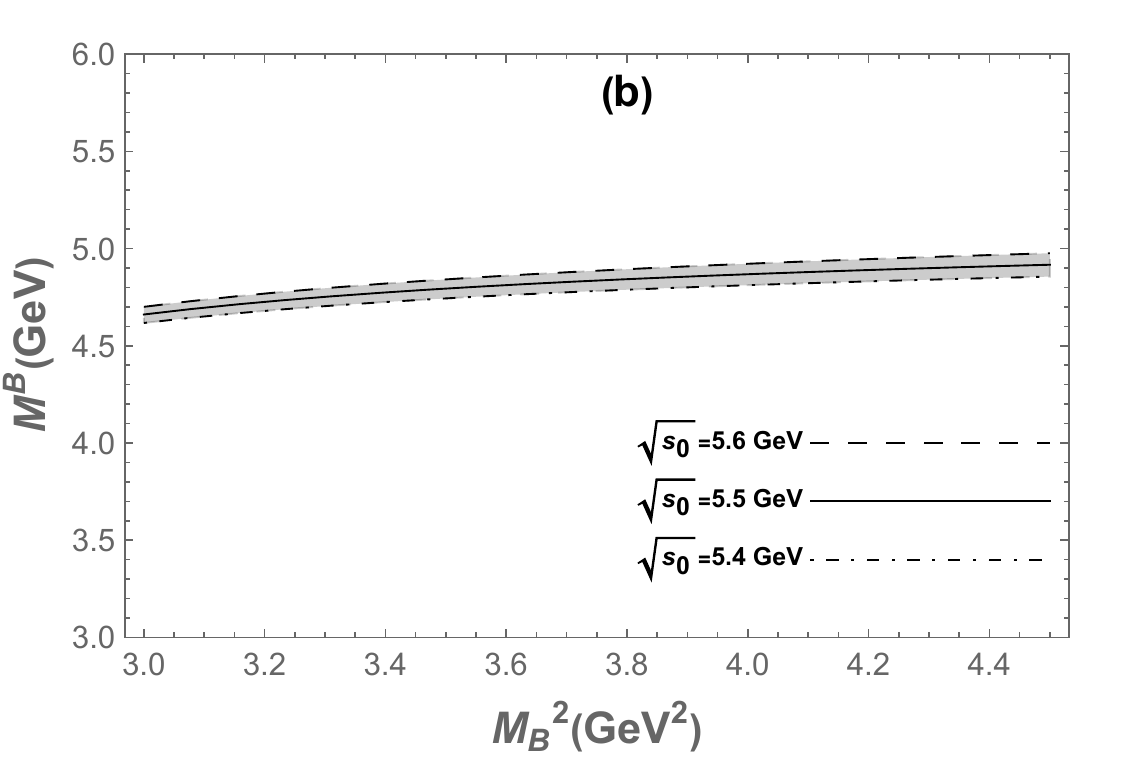}
\caption{The same caption as in Fig \ref{figA}, but for the current in Eq.~(\ref{Jb}).} \label{figB}
\end{figure}

\begin{figure}[h]
\includegraphics[width=6.8cm]{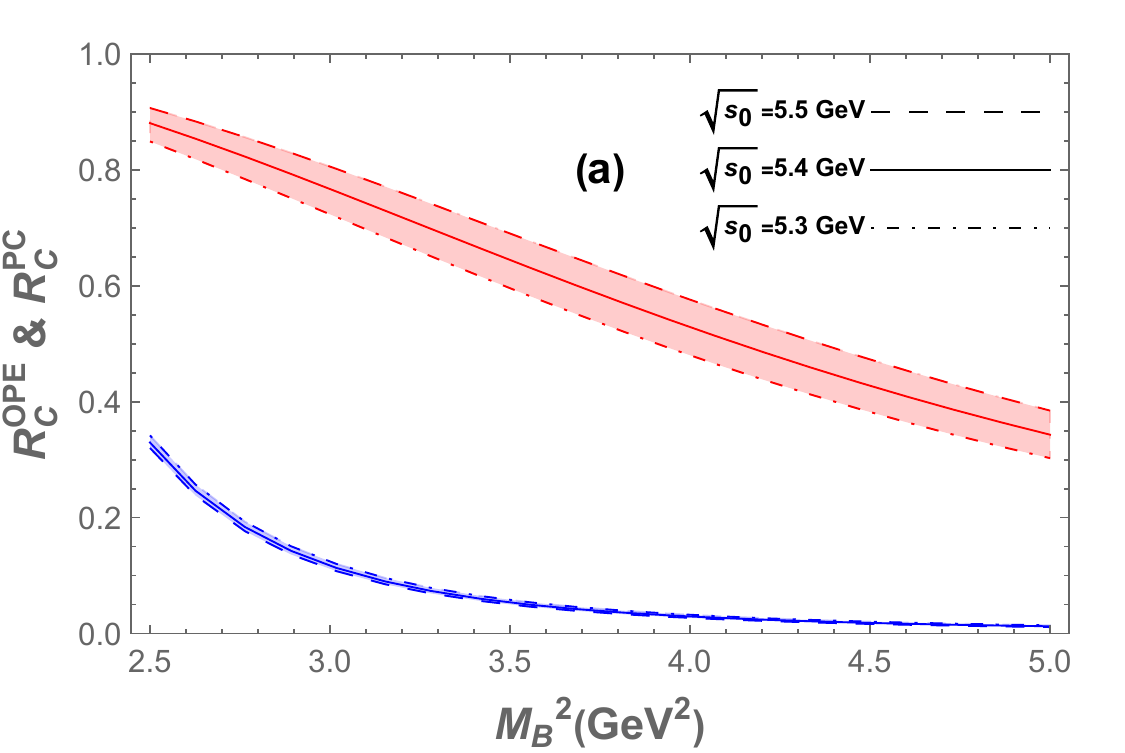}
\includegraphics[width=6.8cm]{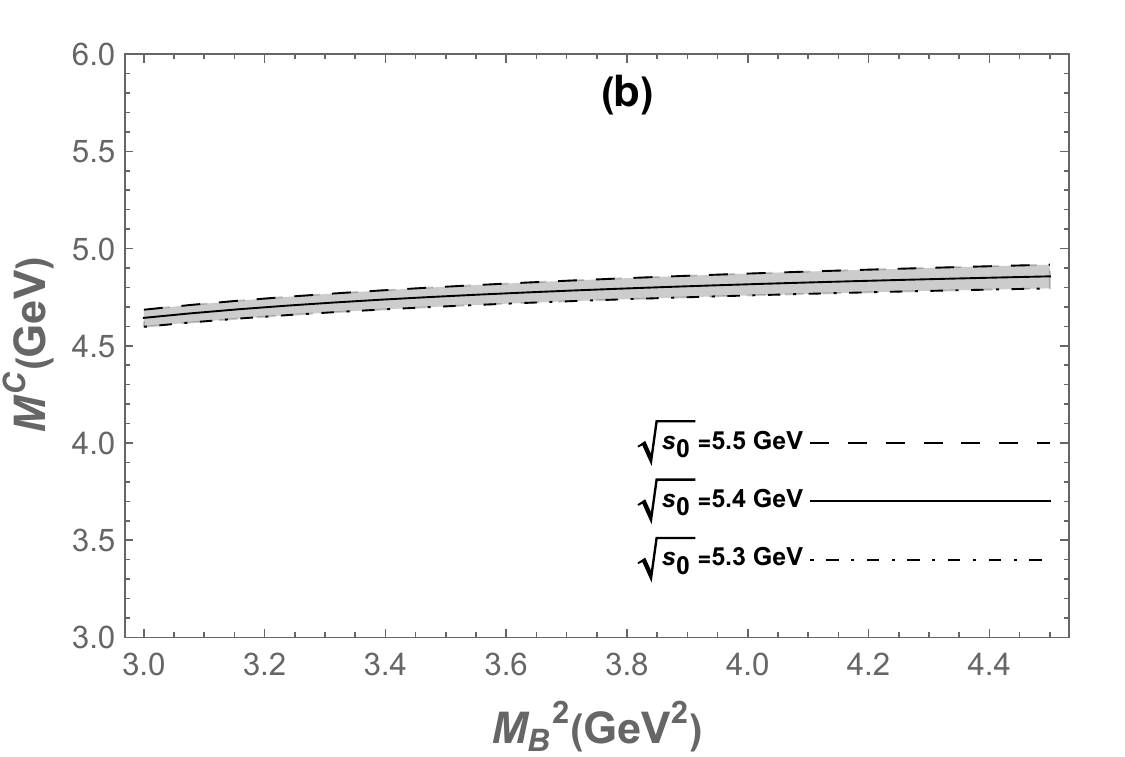}
\caption{The same caption as in Fig \ref{figA}, but for the current in Eq.~(\ref{Jc}).} \label{figC}
\end{figure}

\begin{figure}[h]
\includegraphics[width=6.8cm]{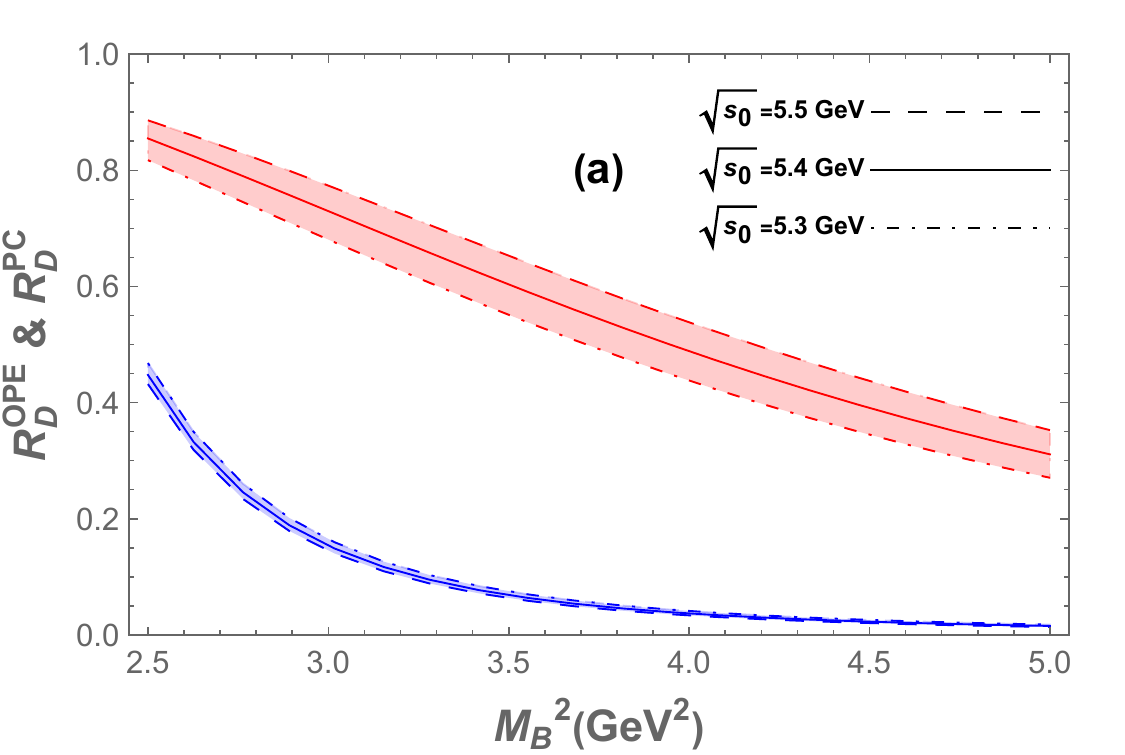}
\includegraphics[width=6.8cm]{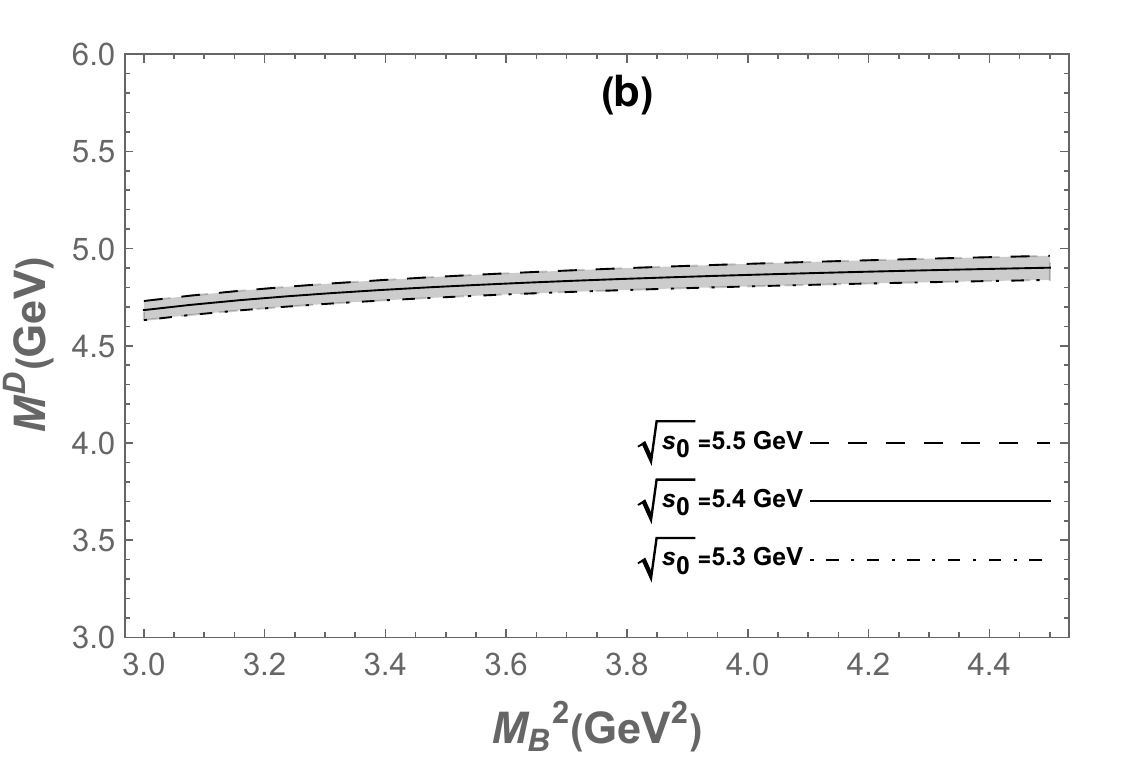}
\caption{The same caption as in Fig \ref{figA}, but for the current  in Eq.~(\ref{Ja}).} \label{figD}
\end{figure}

\begin{figure}[h]
\includegraphics[width=6.8cm]{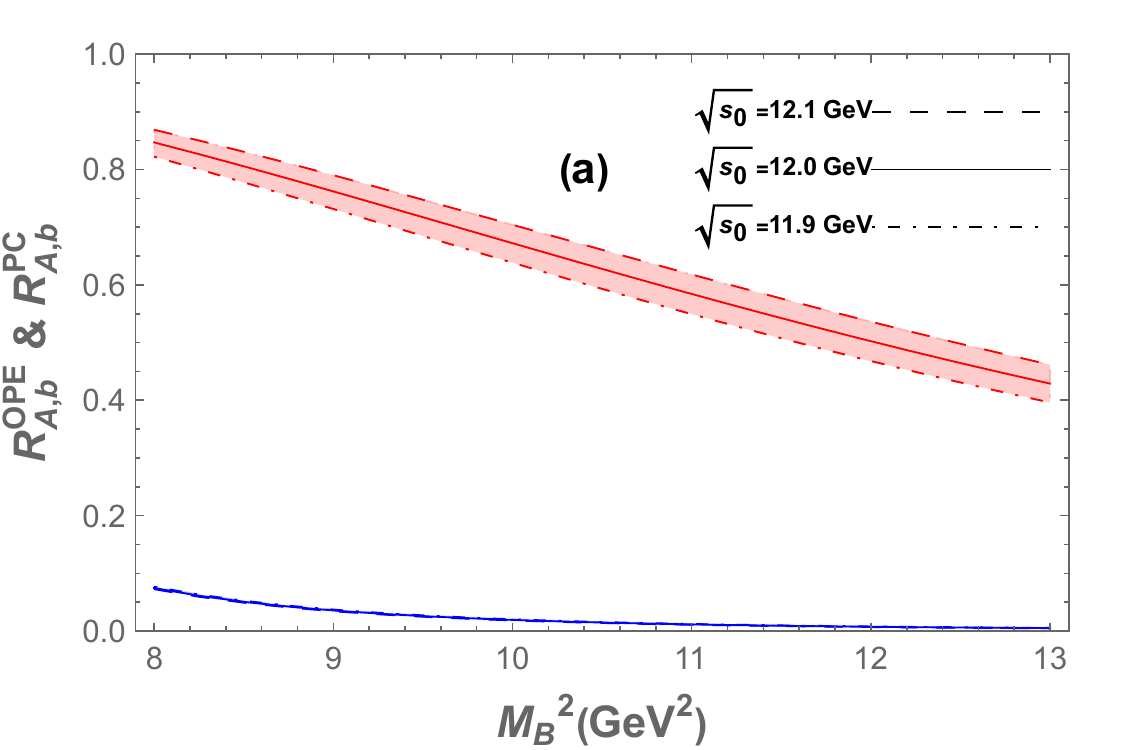}
\includegraphics[width=6.8cm]{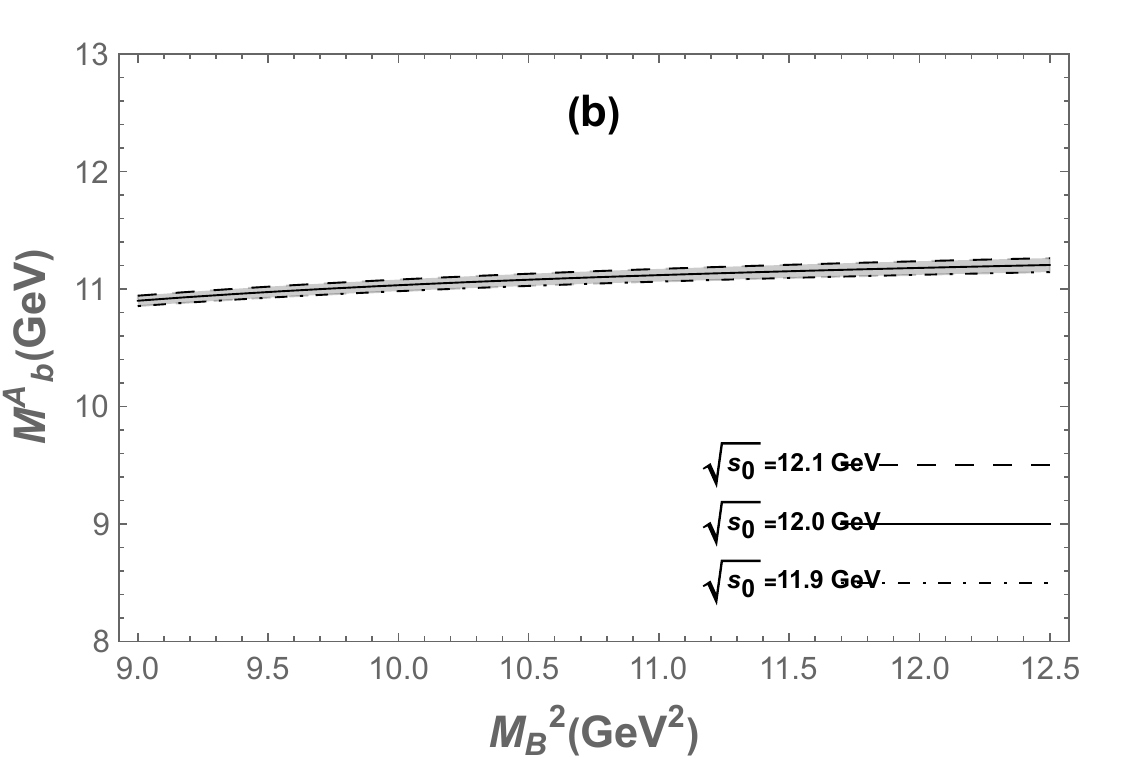}
\caption{The same caption as in Fig \ref{figA}, but for the hidden-bottom tetraquark state.} \label{figAb}
\end{figure}

\begin{figure}[h]
\includegraphics[width=6.8cm]{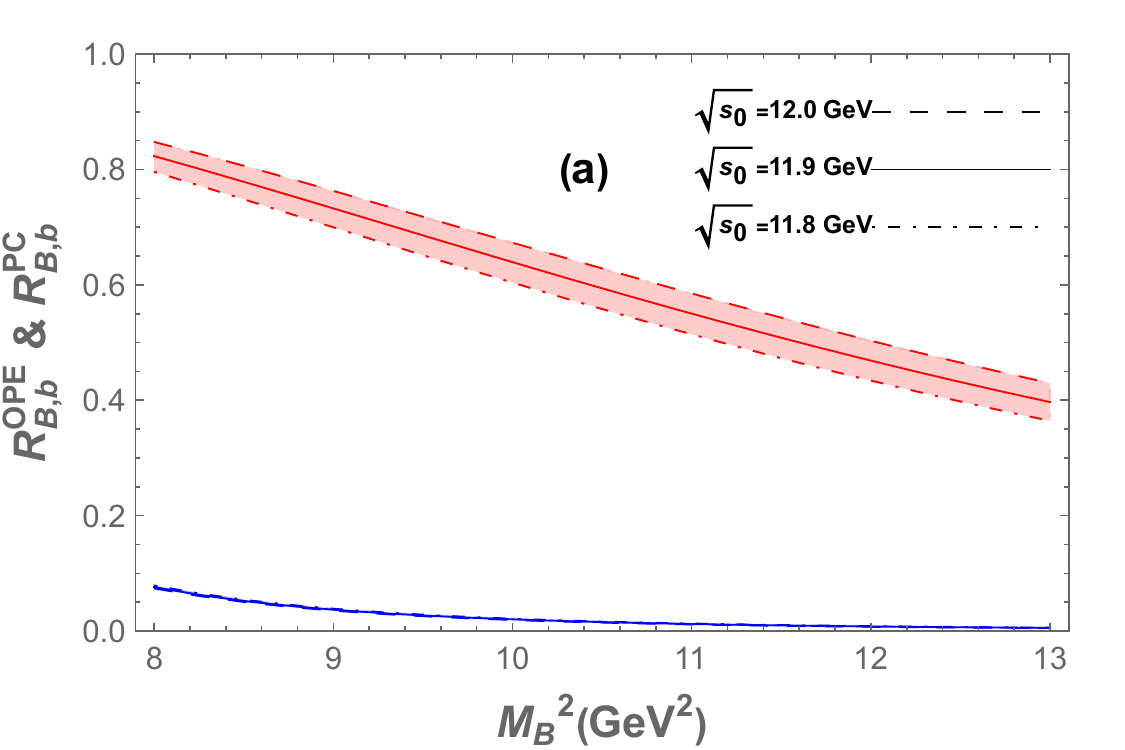}
\includegraphics[width=6.8cm]{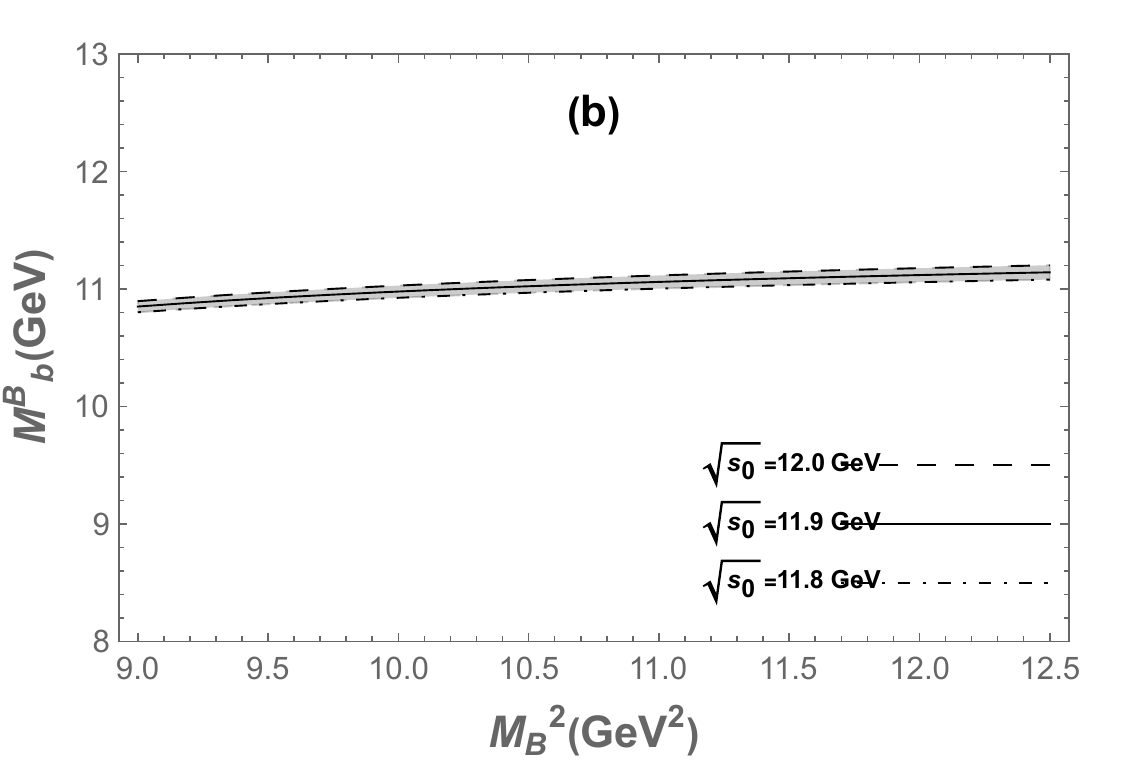}
\caption{The same caption as in Fig \ref{figB}, but for the hidden-bottom tetraquark state.} \label{figBb}
\end{figure}

\begin{figure}[h]
\includegraphics[width=6.8cm]{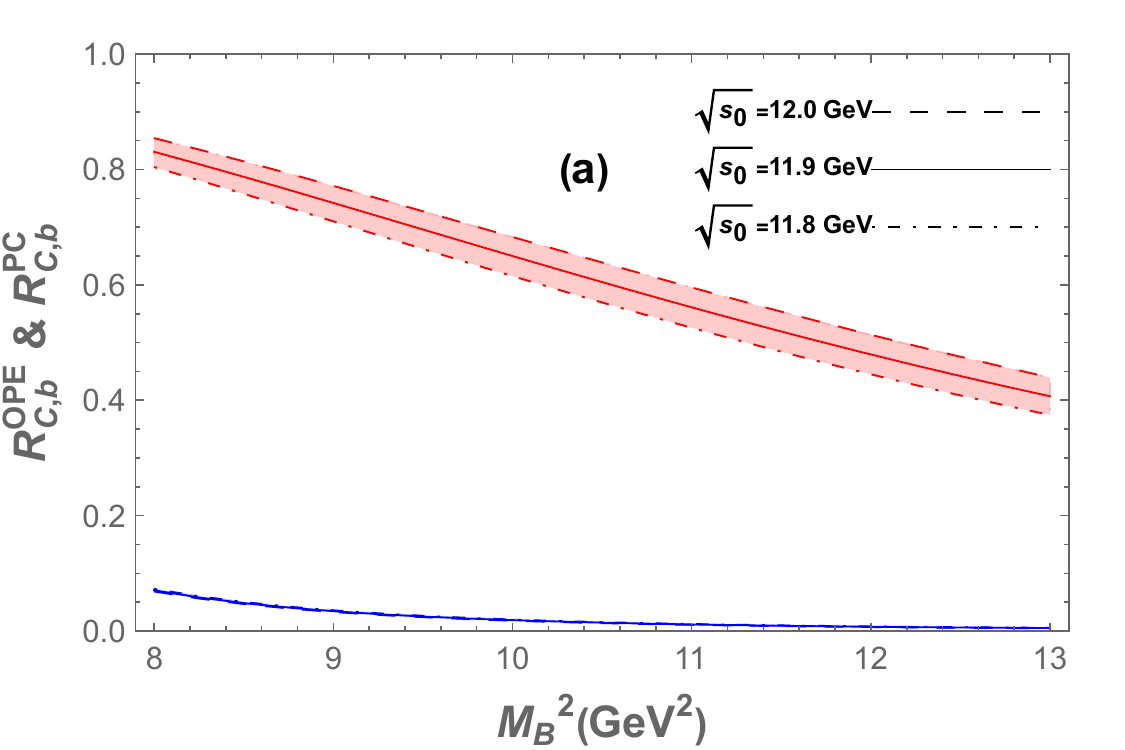}
\includegraphics[width=6.8cm]{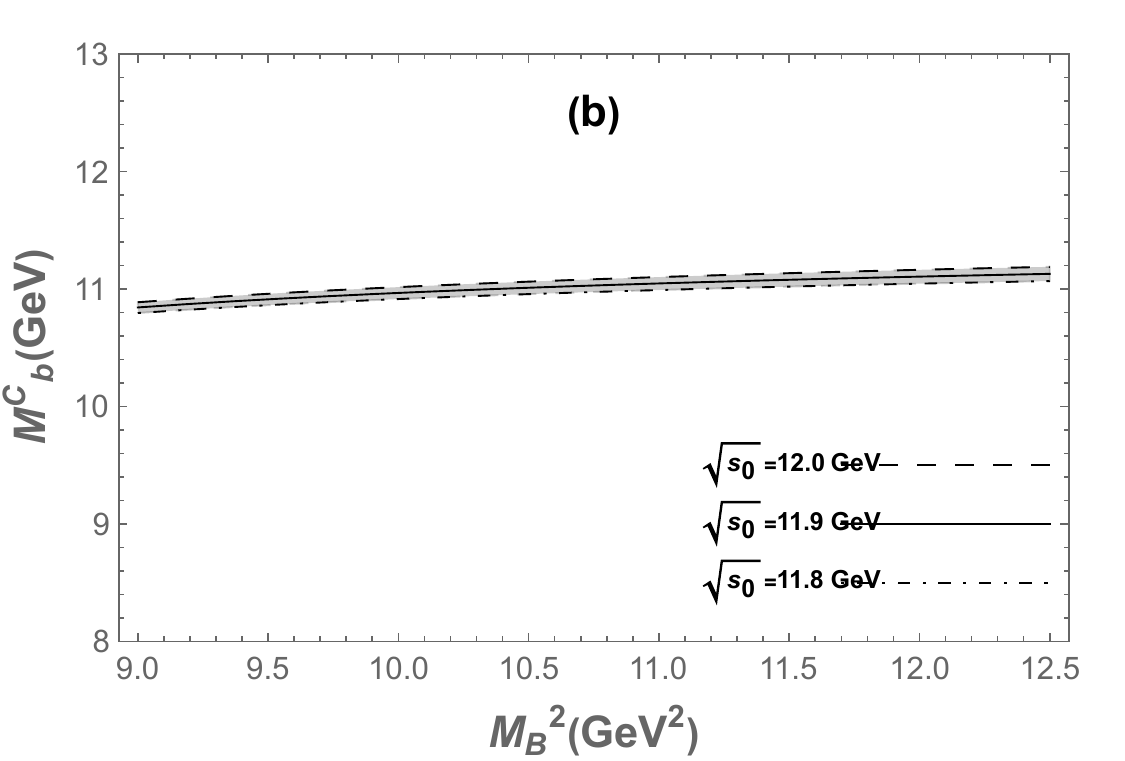}
\caption{The same caption as in Fig \ref{figC}, but for the hidden-bottom tetraquark state.} \label{figCb}
\end{figure}

\begin{figure}[h]
\includegraphics[width=6.8cm]{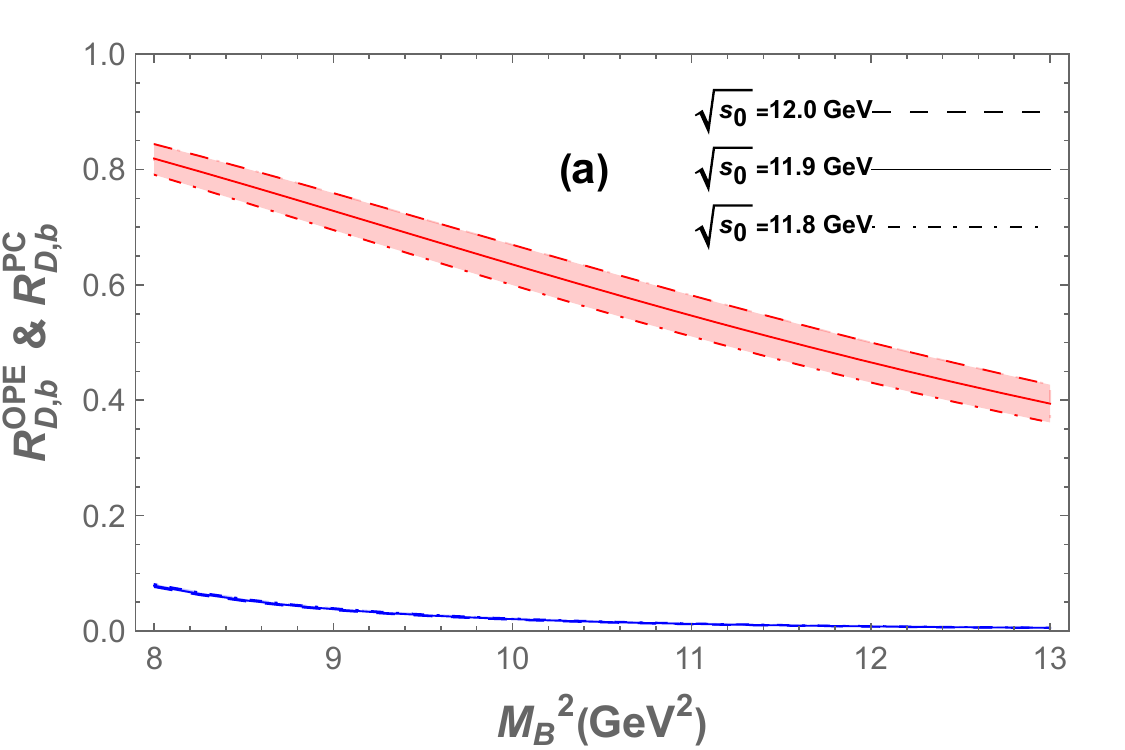}
\includegraphics[width=6.8cm]{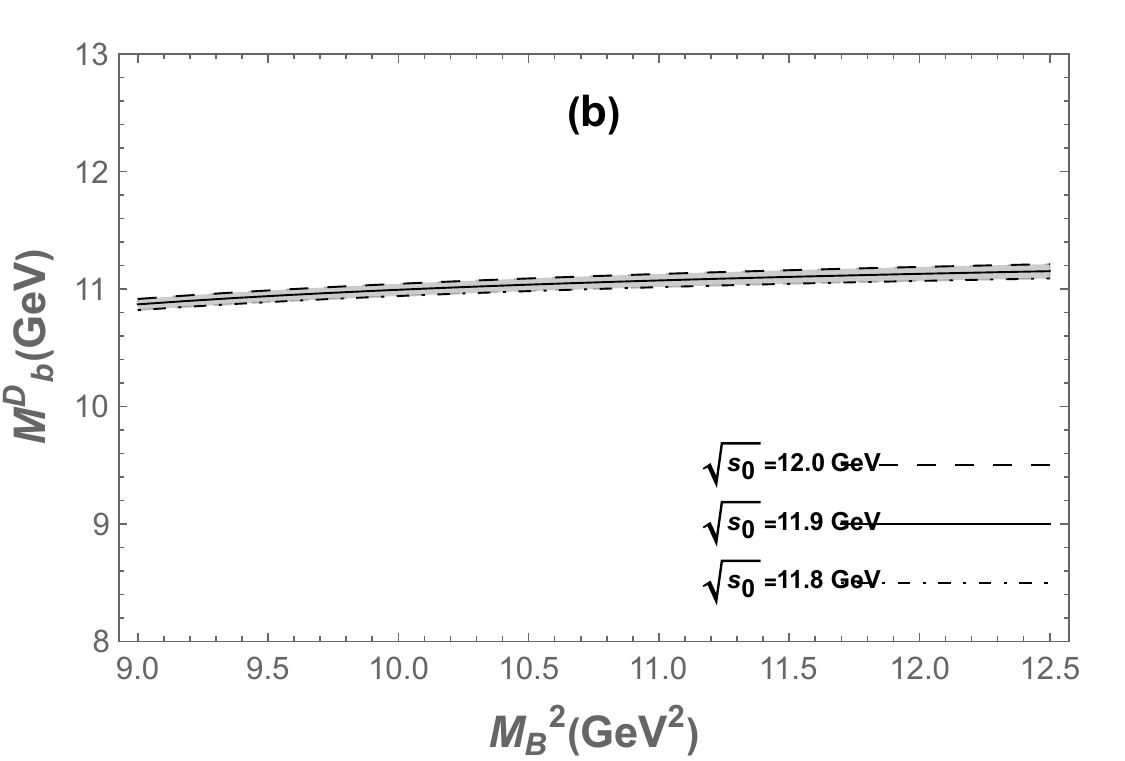}
\caption{The same caption as in Fig \ref{figD}, but for the hidden-bottom tetraquark state.} \label{figDb}
\end{figure}

\end{widetext}
\end{document}